\documentstyle[11pt,twoside,colloqOHP2010,epsfig]{article}
\newcommand{\mearth}{M_\oplus}
\newcommand{\msun}{M_\odot}

\newcommand{\rcapt}{R_{\rm capt}}
\newcommand{\mcore}{M_{\rm core}}

\newcommand{\mz}{M_{\rm Z}}
\newcommand{\mxy}{M_{\rm XY}}

\newcommand{\lsun}{L_{\scriptsize \sun}}
\newcommand{\lj}{L_{\textrm{\scriptsize J}}}
\newcommand{\mj}{M_{\textrm{\scriptsize J}}}
\newcommand{\rj}{R_{\textrm{\scriptsize J}}}
\newcommand{\mstar}{M_*}

\newcommand{\rcore}{R_{\rm core}}

\newcommand{\mdotz}{\dot{M}_{\rm Z}}
\newcommand{\mdotxy}{\dot{M}_{\rm XY}}

\newcommand{\beq}{\begin{equation}}
\newcommand{\eeq}{\end{equation}}
\newcommand{\beqa}{\begin{eqnarray}}
\newcommand{\eeqa}{\end{eqnarray}}

\begin{document}
\title{Theory of planet formation and comparison with observation: \\ Formation of the planetary mass-radius relationship }
\author{Christoph Mordasini$^1$, Yann Alibert$^2$, Willy Benz$^2$, and Hubert Klahr$^1$}
\affil{$^1$ Max Planck Institute for Astronomy, K¬\"onigstuhl 17, D-69117 Heidelberg, Germany  [mordasini@mpia.de]}
 \affil{$^2$ Physikalisches Institut, Sidlerstrasse 5, CH-3012 Bern, Switzerland}

\begin{abstract} The planetary mass-radius diagram is an observational result of central importance to understand planet formation. We present an updated version of our planet formation model based on the core accretion paradigm which allows us to calculate planetary radii and luminosities during the entire formation and evolution of the planets. We first study with it the formation of Jupiter, and compare with previous works. Then we conduct planetary population synthesis calculations to obtain a synthetic mass-radius diagram which we compare with the observed one.  Except for bloated Hot Jupiters which can be explained only with additional mechanisms related to their proximity to the star, we find a good agreement of the general shape of the observed and the synthetic $M-R$ diagram. This shape can be understood with basic concepts of the core accretion model. 
\end{abstract}

\section{Introduction}
The number of known transiting extrasolar planets is increasing rapidly. Combined with radial velocity measurements which yield the mass of the planet, one gets the planetary mass-radius diagram, which is an observational result of similar importance as the semimajor axis-mass diagram. The reason for this is that one can derive the mean density of the planet, which constrains, at least to some extent, the internal structure which is of central importance to understand the nature (Leconte et al. 2009) and, as we shall see, the formation of the planet.  In this work we discuss recent steps towards a synthetic mass-radius diagram as obtained through planetary population synthesis. Keeping in mind the goal of the colloquium, we concentrate the comparison of theoretical and observed results on this subject. A comparison of numerous other theoretical and observational results mostly obtained by the radial-velocity technique can be found in Mordasini et al. (2009b). 

The positive fact which allows transits to be observable with a reasonable probability, namely the large proximity of almost all transiting extrasolar planets to their host star, is in the context of formation modeling in the same time also a complication: The structure of the protoplanetary close to the star is probably much more complex than further out due to the strong effects of stellar irradiation and magnetic fields (e.g. Dullemond \& Monnier 2010). The properties of the protoplanetary disks are however very important for the formation, as they are the boundary conditions for planetary growth and migration. Due to this are many mechanism affecting the formation of close-in planets not yet well understood, like for example stopping mechanisms for migration (e.g. Ogihara et al. 2010). Also the further evolution of close-in planets is more complicated and not yet fully understood, as indicated by the ``radius anomaly'' (Leconte et al., this volume). This complicates the derivation of direct constraints on formation models, and  highlights the importance of the detection of further transiting planets at larger distances, say outside 0.1 AU.
   
Before addressing results obtained by the population synthesis, it is essential to present the basic concepts of the planet formation theory, and how they are described in the formation model.  The reason is that the output of the model is of fundamental importance for the properties of the synthetic planetary population. As most transiting extrasolar planets are still giant planets, it is natural to focus on the formation of this type of planets.

\section{Giant planet formation model}\label{sect:model}
The results presented here were obtained using an updated version of the planet formation code of Alibert, Mordasini, \& Benz (2004), described in details in Alibert et al. (2005). In this version (Mordasini et al. in prep.), we have added the calculation of the planetary structure during the full formation and subsequent evolution of the planet, from a tiny planetary embryo to a billion years old giant planet. This allows to calculate planetary radii (instead of the mass only as in previous versions) which is obviously of primary interest for transit observations. 

The formation model used here relies on the core accretion paradigm. Its basic conception is to follow the concurrent growth of an initially small solid core consisting of ices and rocks, and its surrounding gaseous envelope, embedded in the protoplanetary disk (Perri \& Cameron 1974; Mizuno et al. 1978; Bodenheimer \& Pollack 1986, hereafter BP86). Within the core accretion paradigm, giant planet formation happens as a two step process: first a solid core with a critical mass (of order 10-15 $\mearth$) must form, then the rapid accretion of a massive gaseous envelope sets in. 

\subsection{Core growth}
The growth rate of the solid core is described with a Safronov (1969) type rate equation,
\beq
\frac{d\mz}{dt}=\pi \rcapt^2 \Omega\Sigma_{\rm p}  F_{\rm G}
\eeq
(cf. Lissauer 1993) where $\Omega$ is the Keplerian frequency of the protoplanet at an orbital distance $a$ around the star of mass $\mstar$, $\Omega=\sqrt{G \mstar/a^{3}}$, $\Sigma_{p}$ is the surface density of the field planetesimals, $\rcapt$ is the capture radius of the embryo, and $F_{\rm G}$ is the gravitational focussing factor (Greenzweig \& Lissauer 1992). It reflects the fact that due to gravity, the effective collisional cross section of the body is larger than the purely geometrical one, $\pi \rcapt^{2}$, as the trajectories get bended towards the big body. The capture radius in turn is larger than the core radius $\rcore$ due to the presence of the gaseous envelope.

In previous versions of the formation model, we assumed a constant core density of 3.2 g/cm$^{3}$. In order to obtain radii also for solid dominated lower mass planets, we have replace this assumption with an internal structure model for the core. This model was originally developed for the study of GJ 436b in Figueira et al. (2009). We use the EOS from Fortney, Marley. \& Barnes (2007), and assume a three layer differentiated planet model (iron/nickel, silicates, ices). The silicate-iron/nickel ratio is fixed to 2/3 - 1/3 as found for the Earth, and the ice fraction is given by the formation model (accretion of planetesimals inside or outside the iceline). We include the effect of the external pressure by the envelope, which leads to significant compression of the core for giant planets (Baraffe, Chabrier, \& Barman 2008).

\subsection{Envelope growth}
The gas accretion rate of the planet (at least in the early phases, cf. below) is obtained by solving the one dimensional, hydrostatic planetary structure equations. These equations are similar to those for the interiors of stars, except that the energy release by nuclear fusion is replaced by the heating by impacting planetesimals $\epsilon$, which are an important energy source during the early formation stage. The other equations are the standard equation of mass conservation, of hydrostatic equilibrium, of energy conservation and of energy transfer which are given as (e.g. BP86; Guillot 2005; Broeg 2010): 
\beq\label{eq:structureeqs}
\begin{array}{lr}
\frac{dm}{dr}=4 \pi r^{2} \rho  & \frac{dP}{dr}=-\frac{G m}{r^{2}}\rho \\[5pt]
\frac{dl}{dr}=4 \pi r^{2} \rho \left(\epsilon-T\frac{\partial S}{\partial t}\right) & \frac{dT}{dr}=\frac{T}{P}\frac{dP}{dr}\nabla
\end{array}
\eeq
In these equations, $r$ is the radius as measured from the planetary center, $m$ the mass inside r (including the core mass $\mz$), $l$ the luminosity at $r$, $\rho, P, T, S$ the gas density, pressure, temperature and entropy, $t$ the time, and $\nabla$ is given as
\begin{equation}\label{eq:nabla}
\begin{array}{lr}
\nabla=\frac{d\,{\rm ln}\,T}{d\,{\rm ln}\,P}={\rm min}(\nabla_{\rm ad},\nabla_{\rm rad}) & \nabla_{\rm rad}=\frac{3}{64 \pi \sigma G}\frac{\kappa l P}{T^{4} m}
\end{array}
\end{equation}
i.e. by the minimum of the adiabatic gradient $\nabla_{\rm ad}$ which is directly given by the equation of state  (in convective zones) or the radiative gradient  $\nabla_{\rm rad}$  (in radiative zones) where $\kappa$ is the opacity and $\sigma$ is the Stefan-Boltzmann constant.

\subsubsection{Calculation of the luminosity}
For the planetary population synthesis, where the evolution of thousands of different planets is calculated, we need a stable and rapid method for the numerical solution of these equations. We have therefore replaced the ordinary equation for $dl/dr$ by the assumption that $l$ is constant within the envelope, and that we can derive the total luminosity $L$ (including solid and gas accretion, contraction and release of internal heat) and its temporal evolution by total energy conservation arguments, an approach somewhat similar to Papaloizou \& Nelson (2005). We first recall that $-dE_{\rm tot}/dt =L$ and that in the hydrostatic case, the total energy is given as
\beq
E_{\rm tot}=E_{\rm grav}+E_{\rm int}=-\int_0^M \frac{G m}{r}\ d m +  \int_0^M u\ d m \ \dot{=}-\xi \frac{G M^2}{2 R}
\eeq
where $u$ is the specific internal energy, $M$ the total mass, and $R$ the total radius of the planet. We have defined a parameter $\xi$, which represents the distribution of mass within the planet and its internal energy content. The $\xi$ can be found for any given structure at time $t$ with the equations above. Then one can write
\begin{equation}
\begin{array}{l}
-\frac{d}{dt}E_{\rm tot}=L=L_{M}+L_{R}+L_{\xi}=\frac{\xi G M}{R} \dot{M} \ - \  \frac{\xi G M^2}{2 R^2} \dot{R} \ + \ \frac{G M^2}{2 R}\dot{\xi} 
\end{array}
\end{equation}
where $\dot{M}=\mdotz+\mdotxy$ is the total accretion rate of solids and gas, and $\dot{R}$ is the rate of change of the total radius.  All quantities except $\dot{\xi}$ can readily be calculated at time $t$. We now set 
\begin{eqnarray}
L&\simeq&C\left(L_{M}+L_{R}\right). \label{eq:lumi2}
\end{eqnarray}
The factor $C$ corrects that we neglect the $L_{\xi}$ term, and is obtained in this way: a posteriori, one can calculate the total energy in the new structure at $t+dt$, which gives the exact luminosity as $L_{\rm ex}=-[E_{\rm tot}(t+dt)-E_{\rm tot}(t)]/dt$.  By setting $C=L_{\rm ex}/L$ one obtains the correction factor so that exact energy conservation would have occurred. As an approximation, we then use this $C$ for next time step. One finds that in this way, the estimated luminosity $L$ and the actual luminosity $L_{\rm ex}$ agree generally very well, provided that $dt$ is small enough. Note that the release of energy by the contraction of the core - the dominant contribution from it for giant planets  (Baraffe et al. 2008) - is automatically included in the total luminosity in this method.

We have tested the impact of the simplified luminosity calculation by comparison with calculations based on the full set of structure equations and found good agreement (Broeg et al. in prep, sect. \ref{sect:jupiform}). Regarding the assumption that $dl/dr=0$, we have in tests adopted another prescription namely a linear increase with $m$ from $l(R_{\rm core})=L_{\rm core}=(G \mz/\rcore) \mdotz$ to $l(R)=L$. One finds only very minor differences like for example a variation of the time until runaway gas accretion in the case nJ1 discussed below by about 2\%. The reason for this small effect is the following: The radial variation of the luminosity is only relevant in the radiative zones of the planet (see eq. \ref{eq:nabla}). Significant radiative zones however  only exist during the early phase of the planet's formation  (BP86). But then, a large part of the luminosity is generated by the core luminosity, so our approximation is close to the exact solution. During the long term evolution, the planet is nearly fully convective (e.g. Guillot 2005), so that the radial variation of $l$ is irrelevant, too.  

Finally one must take into account that a part of the released gravitational energy of newly accreted material is not incorporated into the planetary structure but already lost as $L_{\rm ext}$ in the accretion shock on the planet's surface or in the surrounding circumplanetary disk (Bodenheimer, Hubickyj, \& Lissauer 2000; Papaloizou \& Nelson 2005). The luminosity within the planetary structure is then  $L_{\rm int}$ where
\beq
\begin{array}{lr}
L_{\rm ext}=\frac{G M}{R}\mdotxy & L_{\rm int}=L-L_{\rm ext}
\end{array}
\eeq
while an observer would see the total luminosity $L$. It is clear that this treatment is a strong simplification of the exact physics of the accretion shock (e.g. Stahler, Shu, \& Taam 1980).

As one can see, we do not include any mechanism which could lead to the so called ``radius anomaly'' observed for many transiting Hot Jupiter which have radii clearly larger than expected from standard internal structure modeling as done here. The physical mechanisms leading to this bloating are not yet understood, and discussed in dedicated studies (e.g. Leconte et al., this volume).

\subsection{Boundary conditions}\label{subsect:boundary}
In order to solve the structure equations, boundary conditions must be specified, which should also provide a continuous transition between different phases. For the formation and evolution of a giant planet, three different fundamental phases must be distinguished: 

\subsubsection{Attached (or nebular) phase}
At low masses, the envelope of the protoplanet is attached continuously to the background nebula, and the conditions at the surface of the planet  are the pressure $P_{\rm neb}$ and approximately the temperature $T_{\rm neb}$ in the surrounding disk. The total radius $R$  is given in this regime by roughly the minimum of the Hill radius $R_{\rm H}$ and the accretion radius $R_{\rm A}$. The gas accretion rate is found by the solution of the structure equations and physically given by the ability of the envelope to radiate away energy so that it can contract (i.e. its Kelvin-Helmholtz timescale), so that new gas can stream in (Pollack et al. 1996). In equations, we use (Bodenheimer et al. 2000; Papaloizou \& Terquem 1999)
\beq
\begin{array}{ll}
R= \frac{R_{\rm A}}{1 + R_{\rm A}/R_{\rm H}} & P=P_{\rm neb}\\[4pt]
\tau={\rm max}(\rho_{\rm neb}\kappa_{\rm neb}R, 2/3) & T_{\rm int}^{4}=\frac{3 \tau L_{\rm int}}{8 \pi \sigma R^2}\\[4pt]
  T^{4}=T_{\rm neb}^{4}+T_{\rm int}^{4} & l(R)=L_{\rm int}.
  \end{array}
\eeq

\subsubsection{Detached (or transition) phase}
When the gas accretion obtained in this way becomes too high in comparison with the externally possible gas supply, the planet enters the second phase and contracts to a radius which is much smaller than the Hill sphere radius. This is the detached regime of high mass, runaway gas accretion planets. The planet now adjusts its radius to the boundary conditions that are given by an accretion shock on the surface for matter falling with free fall velocity $v_{\rm ff}$ onto the planet from the Hill sphere radius, or probably more realistically, by conditions appropriate for the interface to a circumplanetary disk (Papaloizou \& Nelson 2005). 

In this phase, the gas accretion rate $\mdotxy$ is no more controlled by the planetary structure itself, but by how much gas is supplied by the disk and can bypass the gap formed by the planet in the protoplanetary disk, $\dot{M}_{\rm disk}$ (Lubow, Seibert, \& Artymowicz 1999). The boundary conditions are now
\beq
\begin{array}{ll}
  \mdotxy=\dot{M}_{\rm disk}&   v_{\rm ff}=\left[2 G M \left(1/R-1/R_{\rm H}\right)\right]^{1/2}\\[4pt]
    P=P_{\rm neb}+\frac{\mdotxy}{4 \pi R^{2}} v_{\rm ff}+\frac{2 g}{3 \kappa} & \tau={\rm max}(\rho_{\rm neb}\kappa_{\rm neb}R, 2/3)\\ [4pt]
    T_{\rm int}^{4}=\frac{3 \tau L_{\rm int}}{8 \pi \sigma R^2} & T^{4}=(1-A) T_{\rm neb}^{4}+T_{\rm int}^{4}
    \end{array}
\eeq
where $A$ is the albedo (assumed to be the same as for Jupiter) and $g=G M/R^{2}$, and for the luminosity we still have $l(R)=L_{\rm int}$.

\subsubsection{Evolutionary (or isolated) phase}
The last phase starts when the gaseous disk disappears so that the planet evolves at constant mass (we neglect evaporation, as  the minimal allowed semimajor axis $a$ is 0.1 AU in the synthesis). During this phase, we use simple standard stellar boundary conditions in the Eddington approximation and write (e.g. Chandrasekhar 1939) 
\beq
\begin{array}{ll} 
    P=\frac{2 g}{3 \kappa} & T_{\rm int}^{4}=\frac{ L_{\rm int}}{4 \pi \sigma R^2}\\[4pt]
  T_{\rm equi}=280\,{\rm K}  \left(\frac{a}{1 {\rm AU}}\right)^{-\frac{1}{2}}\left(\frac{\mstar}{\msun}\right) &  T^{4}=(1-A) T_{\rm equi}^{4}+T_{\rm int}^{4}
    \end{array}
\eeq
and $l(R)=L_{\rm int}$ which is now also equal the total luminosity $L$. It is clear that the long term evolution is described in a much simpler way than in the Burrows et al. (1997, hereafter BM97) or Baraffe et al. (2003, hereafter BC03) models which employ proper non-gray atmospheres (see Chabrier \& Baraffe 1997).  We have however found that the general agreement in terms of total luminosity or radius is good (cf. sections \ref{sect:jupiform}  and  \ref{sect:lumievo}), with discrepancies of about a factor two in $L$ and $\leq10\%$ in $R$,  sufficient for our purpose of population synthesis. 

\section{Formation of Jupiter}\label{sect:jupiform}
\begin{figure*}
\begin{minipage}{0.5\textwidth}
	      \centering
       \includegraphics[width=\textwidth]{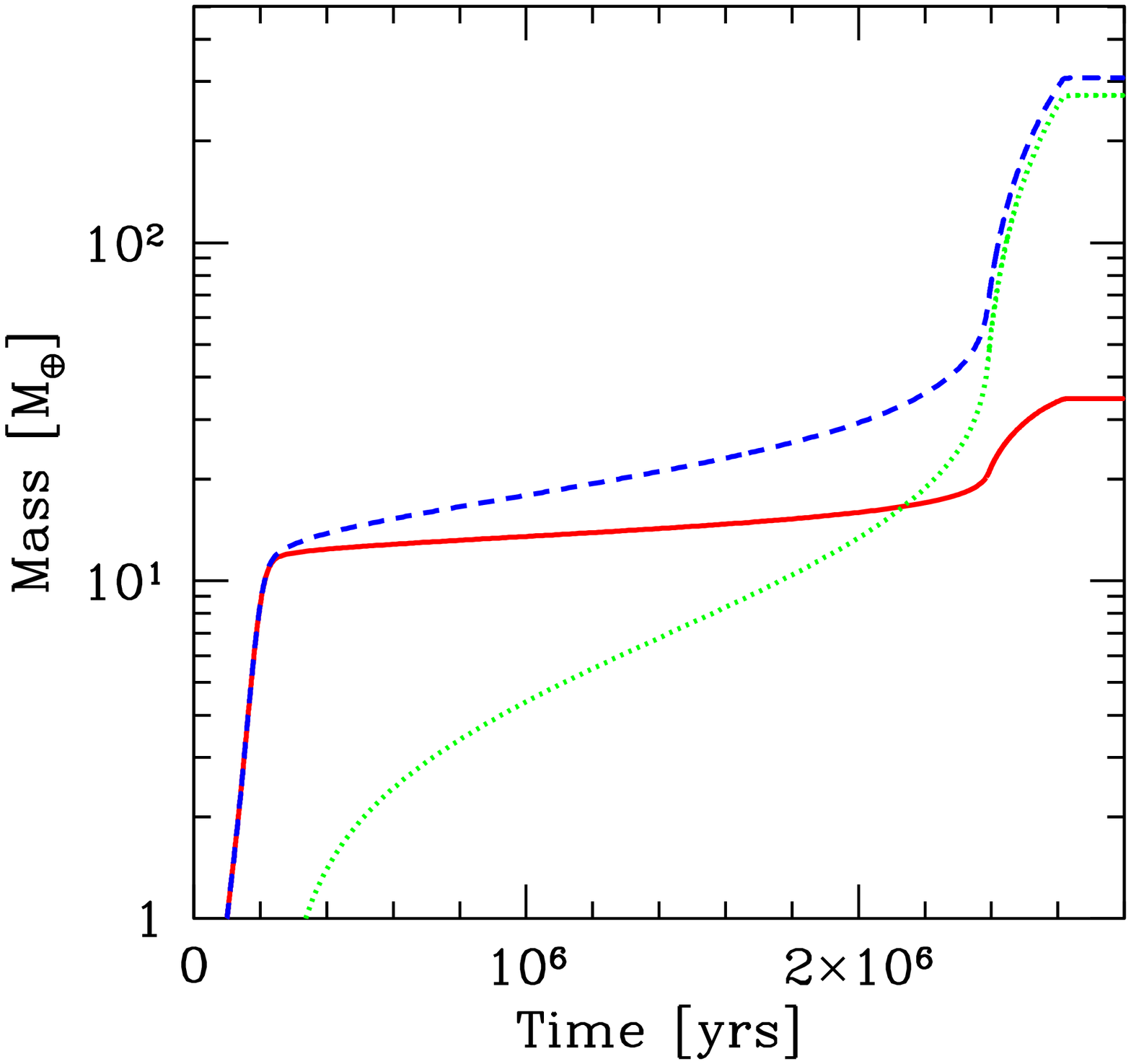}
     \end{minipage}\hfill
     \begin{minipage}{0.5\textwidth}
      \centering
       \includegraphics[width=\textwidth]{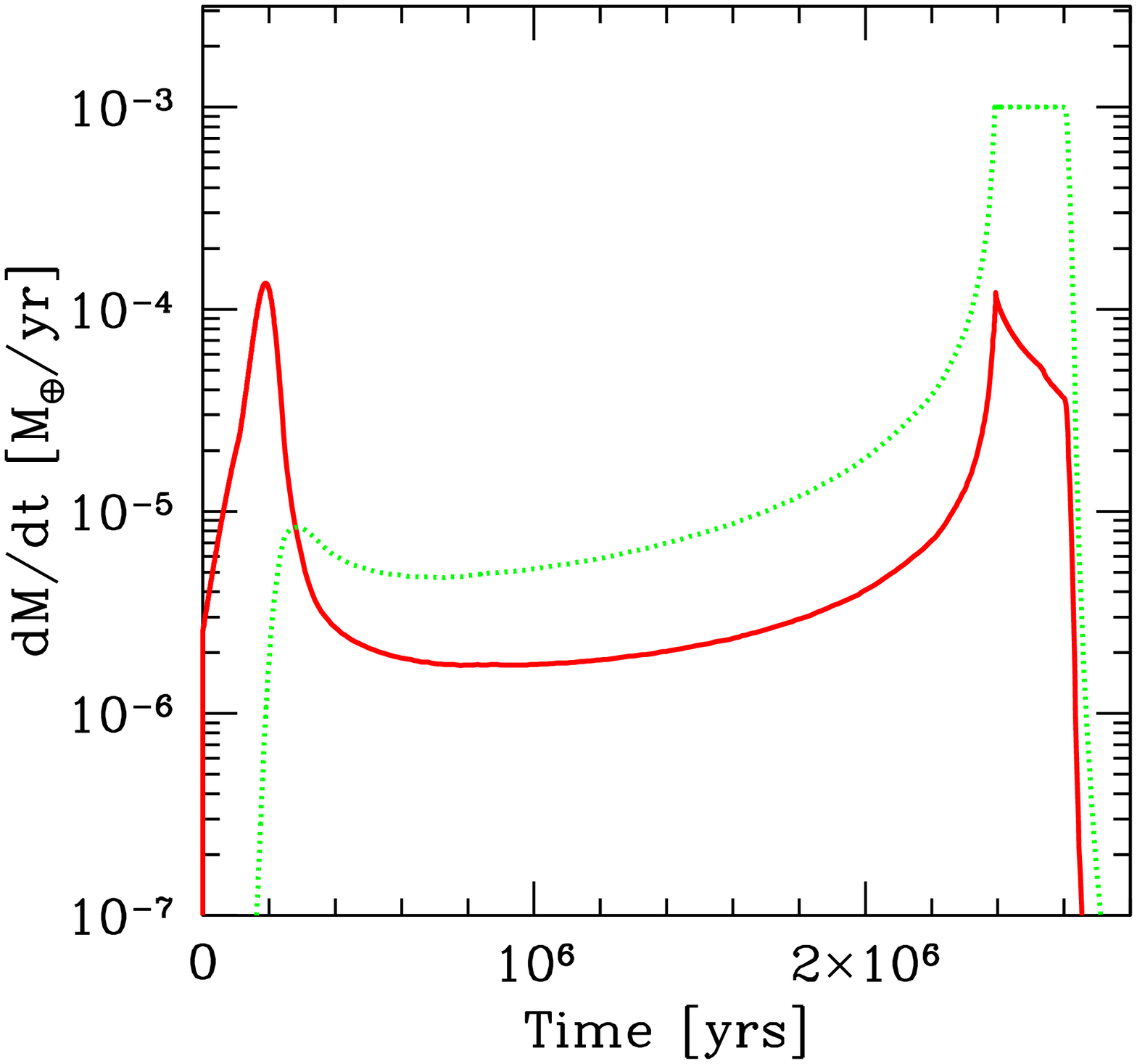}
     \end{minipage}
     \begin{minipage}{0.5\textwidth}
	      \centering
       \includegraphics[width=\textwidth]{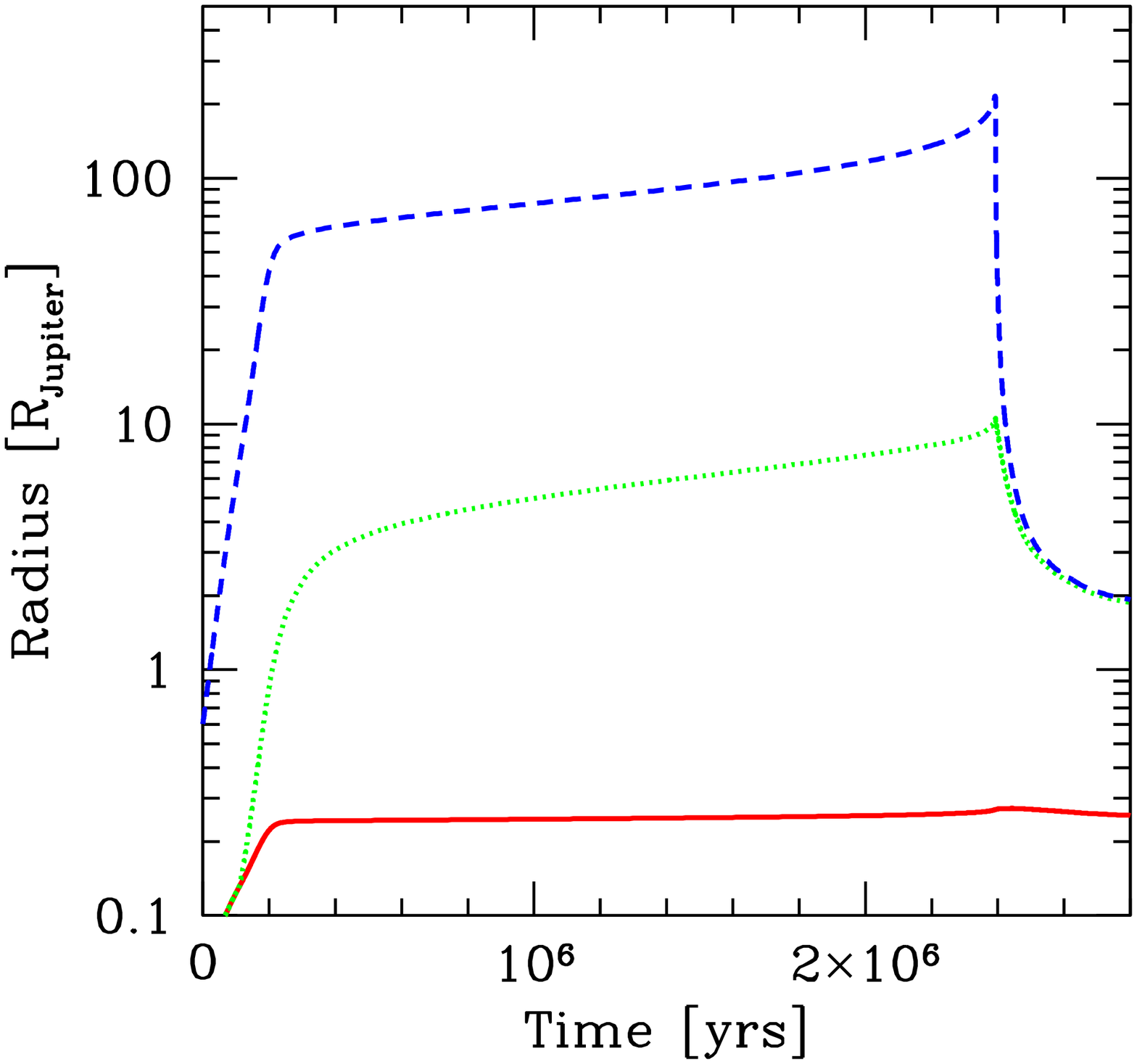}
     \end{minipage}\hfill
     \begin{minipage}{0.5\textwidth}
      \centering
       \includegraphics[width=\textwidth]{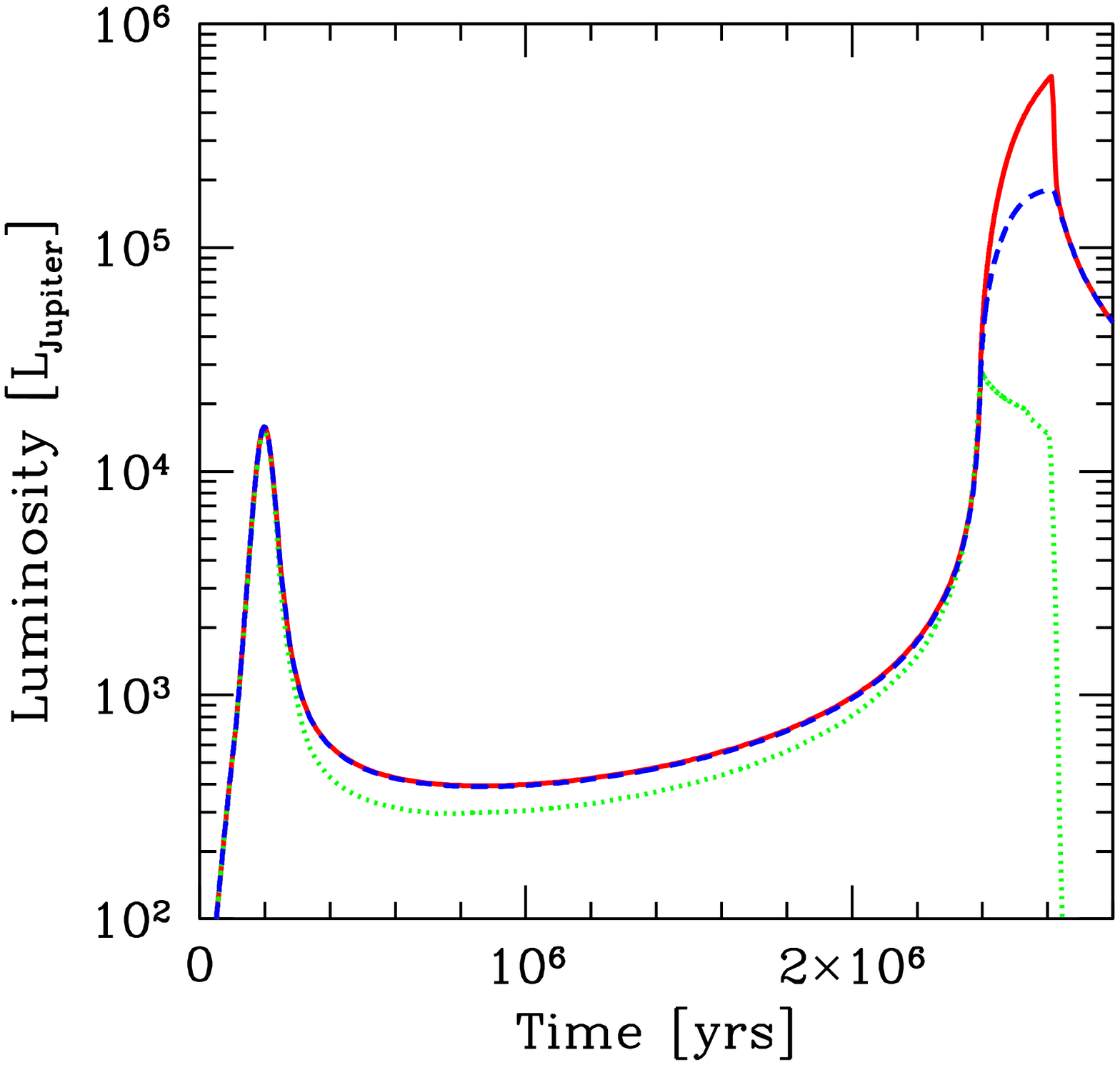}
     \end{minipage}
     \caption{Simulation of the in-situ formation of Jupiter in the nJ6 case. The top left panel shows the evolution of the core mass (red solid line), the envelope mass (green dotted line) and the total mass (blue solid line). The top right panels shows the accretion rate of solids $\mdotz$ (red solid line) and of gas $\mdotxy$ (green dotted line). The limiting gas accretion rate is fixed to $10^{-3} \mearth$/yr. The bottom left panel shows the evolution of the core radius $\rcore$ (red solid line), the total radius $R$ (blue dashed line) and the capture radius $\rcapt$ (green dotted line). The outer radius is initially (during the attached regime) very large, as it is approximately equal to the Hill sphere radius. At about 2.4 Myrs, when the limiting accretion rate is hit, it detaches from the nebula and collapses to a radius of initially about 2 Jupiter radii. The bottom right panel shows the luminosity of the planet in present day intrinsic luminosity of Jupiter ($\lj$=8.7$\times$$10^{-10}\lsun$). The red solid line is the total luminosity $L$, the blue dashed line is the internal luminosity $L_{\rm int}$ and the green dotted line is the core luminosity $L_{\rm core}$. The first peak in the curve is due to the rapid accretion of the core, and the second to the combined effects of runaway gas accretion and envelope collapse.}\label{mordasinifig:coreaccretion} 
\end{figure*}

The formation of Jupiter is of particular importance as a benchmark for all giant planet formation models. This is because for no other giant planet an equally high number of detailed observational constraints exist. Figure \ref{mordasinifig:coreaccretion} shows the formation of Jupiter in our baseline formation simulation (nJ6), using the equations presented in the previous section.

The initial conditions for this simulation mimic J6 in P96 which in particular means that the initial planetesimal surface density $\Sigma_{\rm p}$ is 10 g/cm$^{3}$, and that the grain opacities are  2 \% of the interstellar value. It is clear that the two simulation still cannot yield exactly identical results, as they differ in some other aspects, like the  variable core density and the inclusion of planetesimal ejection in this work, or a different equation of state. 

The simulation shown here is strongly simplified in comparison to the calculations used in the synthesis: In the full model the onset of limited gas accretion (and thus the detachment) as well as the final mass of the planet results in a self-consistent way from the evolution of the gaseous disk (Alibert et al. 2005). Here, the evolution of the disk is switched off. Instead, we fix the maximal allowed $\mdotxy$ to $10^{-3} \mearth$/yr, a value inspired by what is typically seen in the synthesis, and artificially switch off accretion on a short timescale when the mass of the planet approaches 1 Jupiter mass. Also migration is completely switched off, which is otherwise a central component of the formation model (Alibert et al. 2005) as it can lead to much more complex accretion histories (Mordasini, Alibert, \& Benz 2009a).  But in this way we allow for direct comparison with previous works as P96, Hubickyj, Bodenheimer, \& Lissauer (2005) and Lissauer et al. (2009).

The top left and right panel show the characteristic three stages of such in-situ calculations (but all three occur within the attached phase described in sect. \ref{subsect:boundary}): First a rapid build-up of the core until the isolation mass is reached, then a plateau phase characterized by a slow increase of the envelope mass which allows further core growth, and then the transition to gas runaway accretion. This transition is due to the fact that the radiative losses increase, or equivalently that the Kelvin-Helmholtz time scale  $\tau_{\rm KH}$ decreases with  increasing mass.
 
The cross-over point, i.e. the moment when the core and the envelope mass are identical and equal to $M_{\rm cr}$ is found to occur at $t_{\rm cr}=2.14$ Myrs at $M_{\rm cr}=16.6\mearth$. This is in very good  agreement with P96 with $t_{\rm cr}$=$2.75$ Myrs, $M_{\rm cr}$=$16.2\mearth$ or Hubickyj et al. (2005)  who have $t_{\rm cr}$=2.22 Myrs and $M_{\rm cr}$=$16.1\mearth$.

Shortly after the cross-over point, the gas accretion rate increase strongly as gas runaway accretion starts, so that it hits the maximal allowed value of  $10^{-3} \mearth$/yr at t=2.39 Myrs. The total mass of the planet is at this moment is  65.34 $\mearth$, and $\tau_{\rm KH}$ is as short as $\sim2300$ yrs. This moment is also when the second detached phase begins and the collapse of the envelope starts (cf. Lissauer et al. 2009), which is in fact a rapid, but still hydrostatic contraction (BP86). About $2\times 10^{5}$ yrs after this moment, at t=2.61 Myrs, the second luminosity maximum occurs where $L=5.81$$\times$$10^{5}$ $\lj$ ($\log(L/\lsun)=-3.29$). This value is comparable to what is found by Lissauer et al. (2009) for cases which use a similar limiting gas accretion rate. The maximum occurs just before the gas accretion rate is (artificially) reduced down to zero because the mass of the planet approaches 1 Jupiter mass. The planet has then already contracted to a radius of 2.38 $\rj$, and nearly reached its final mass ($M=300.6\mearth$). The highest temperature at the core-envelope interface occurs rough 13000 years after the second luminosity peak, with a $T_{\rm cent}$ of about 66900 K. Then it starts to fall.

\begin{figure*}
\begin{minipage}{0.53\textwidth}
	      \centering
       \includegraphics[width=\textwidth]{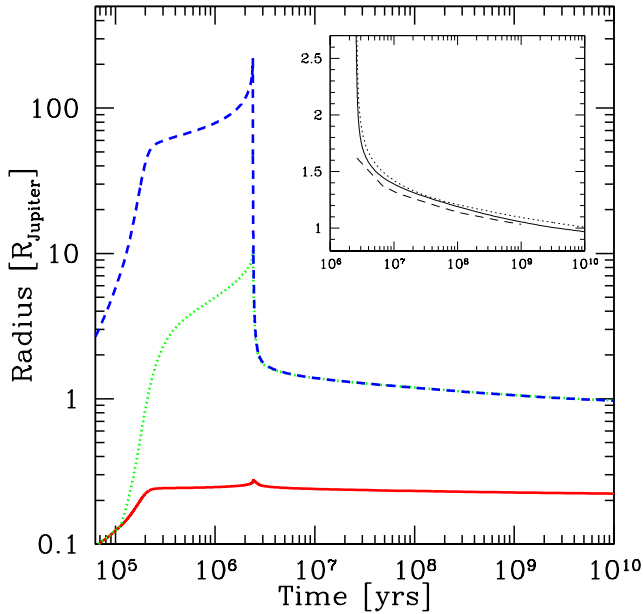}
     \end{minipage}\hfill
     \begin{minipage}{0.4\textwidth}
      \centering
       \caption{Radius $R$ (dashed), core radius $\rcore$ (solid) and capture radius $\rcapt$ (dotted) as a function of time for the nJ6 case. The mass of the core increases during the runaway gas accretion phase by about 15 $\mearth$, but the concurrently increasing pressure on its surface causes the core radius to be almost constant.  The inset figure is a zoom-in onto the late evolution and shows the radius as found in this work (solid line), in BC03 (dashed) and BM97 (dotted).  Note that the different assumed internal compositions have a certain influence on $R$. }\label{mordasinifig:J6rad} 
     \end{minipage}
\end{figure*}
\begin{figure*}
\begin{minipage}{0.65\textwidth}
	      \centering
       \includegraphics[width=\textwidth]{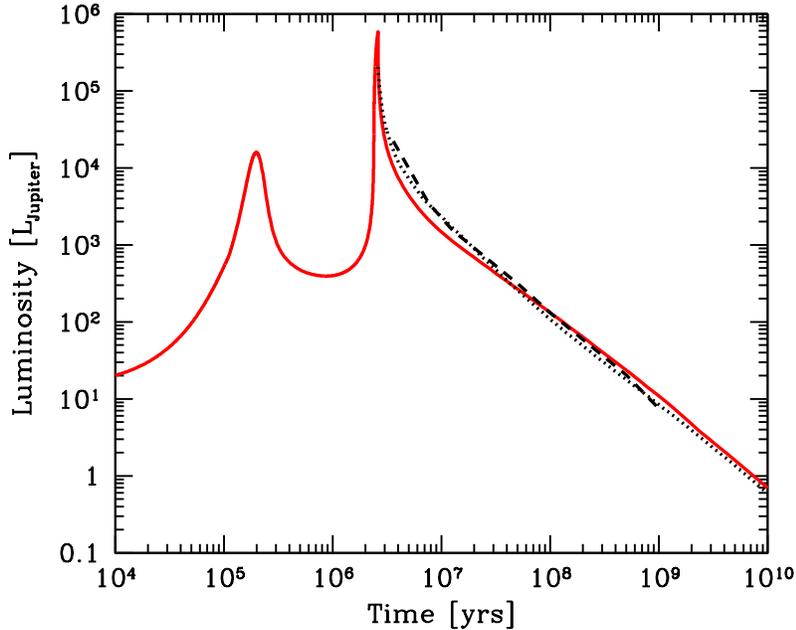}     
       \end{minipage}\hfill
     \begin{minipage}{0.34\textwidth}
      \centering
       \caption{The total luminosity $L$ as a function of time for the complete evolution of the nJ6 case (solid line). During the second maximum, occurring at $t=2.61$ Myrs, a luminosity of about $5.81$$\times$$10^{5}$ $\lj$ equal to $\log(L/\lsun)=-3.29$ is attained.  The dashed line shows BC03 and the dotted one is BM97.}\label{fig:J6lumi} 
     \end{minipage}
\end{figure*}

Figure \ref{mordasinifig:J6rad} and \ref{fig:J6lumi} show the  radius and luminosity of the planet as a function of time, including the long term evolution over Gigayears when the mass is constant.  The evolution now occurs slowly by gradual contraction and cooling. For comparison, the temporal evolution of the radius and the luminosity of a 1 $\mj$ planet as found by BC03 and BM97 is also shown. One finds good agreement. 

The planet has the following key properties after 4.6 Gyrs: A total mass of 307.3 $\mearth$, a heavy element mass $\mz=34.6\mearth$ (Z=0.11) which falls well inside the (wide) range of possible values for Jupiter of 10 to 42 $\mearth$ (Guillot \& Gautier 2007), a radius of 0.99 $\rj$, an intrinsic luminosity of $1.78 \lj$, and a surface temperature $T=133$ K (measured value 124 K, Guillot \& Gautier 2007). The temperature at the core-envelope interface is about 17300 K, similar to what is found in more complex models (Guillot \& Gautier 2007). We thus see that the model, in particular with the simplifications we made, is able to fulfill the most important observational constraints coming from Jupiter.

\begin{figure*}
     \begin{minipage}{0.5\textwidth}
	      \centering
       \includegraphics[width=\textwidth]{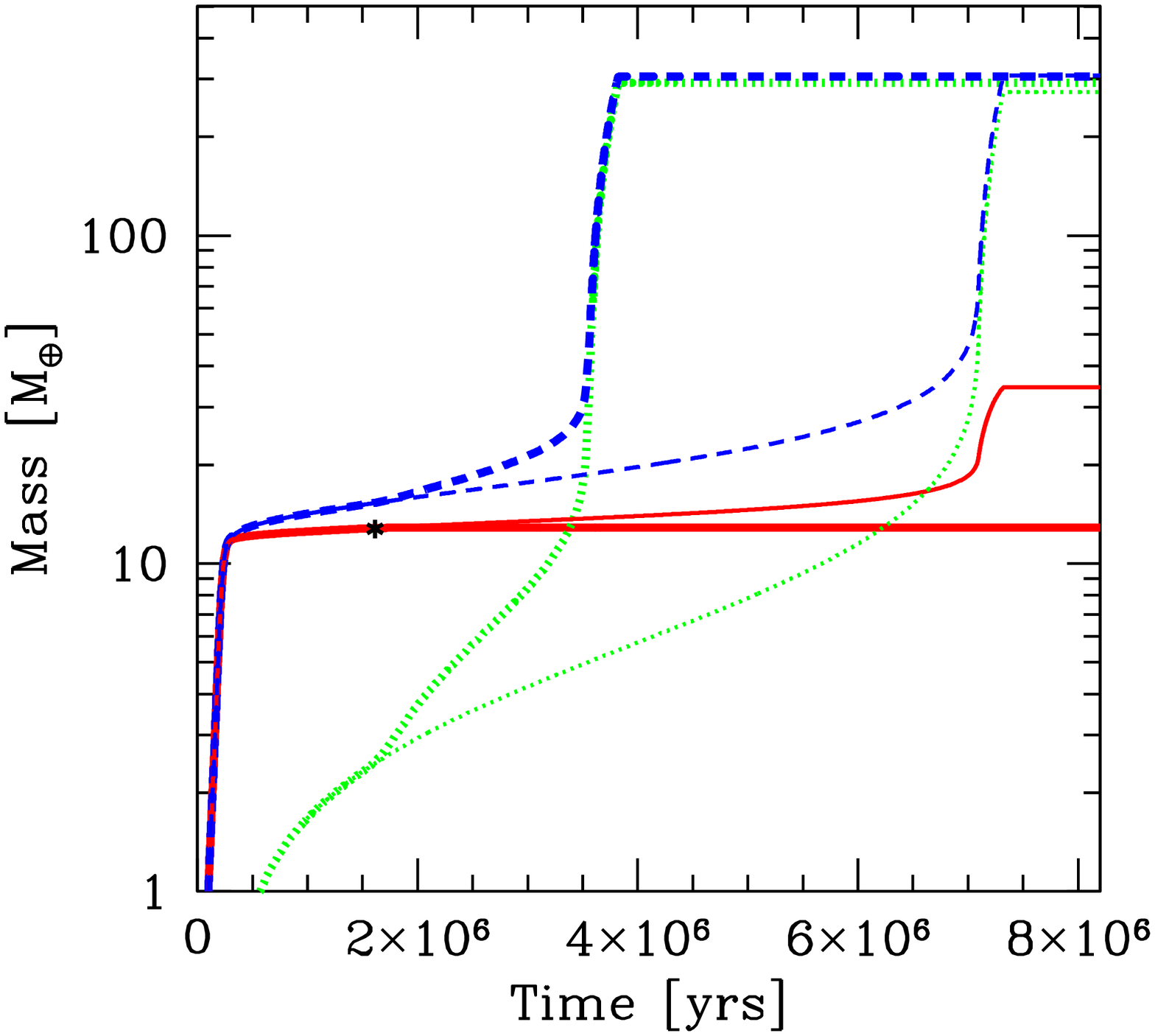}
     \end{minipage}\hfill
     \begin{minipage}{0.5\textwidth}
      \centering
       \includegraphics[width=\textwidth]{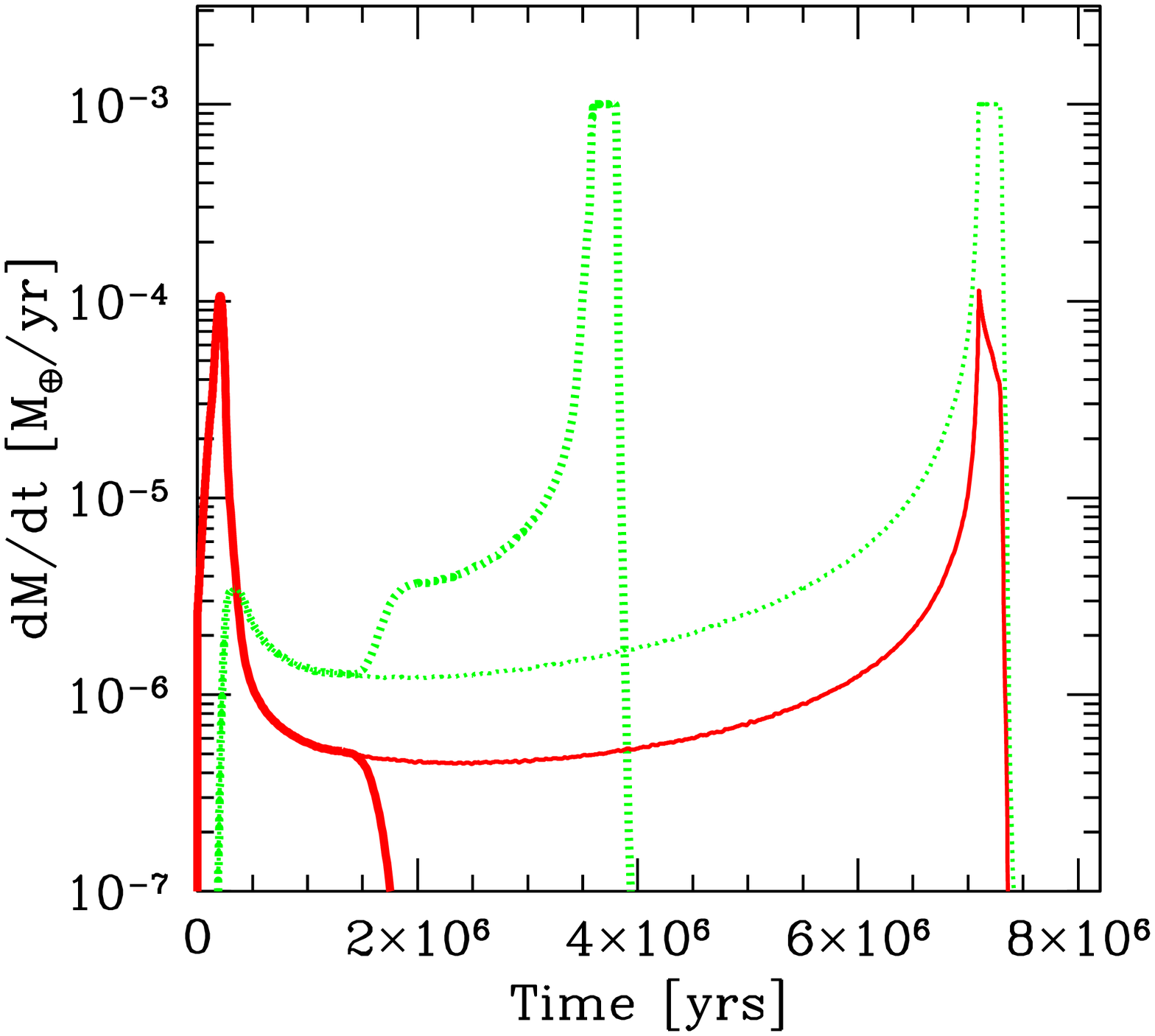}
     \end{minipage}
    \caption{Left panel: Core, envelope and total mass for the cases nJ1a (thick lines) and nJ1 (thin lines). The lines have the same meaning as in Fig. \ref{mordasinifig:coreaccretion}, top left panel. The black symbol indicates the moment in time when for nJ1a $\mdotz$ is artificially ramped down. Right panel: Corresponding gas and solid accretion rates. The meaning of the curves corresponds to the top right panel of Fig. \ref{mordasinifig:coreaccretion}.}\label{fig:J1J1a} 
    \end{figure*}

Figure \ref{fig:J1J1a} shows the mass and mass accretion rate as a function of time for two other Jupiter formation calculations. The thin lines (case nJ1) correspond to a set-up equivalent to J1 in P96, which means that it only differs from nJ6 by the assumption of full grain opacities in the envelope. The higher opacity has the well known effect (e.g. BP86; Ikoma, Nakazawa, \& Emori 2000) of increasing the formation time. Indeed, the cross-over time now occurs at $t=6.62$ Myrs (P96: 7.58 Myrs;  Hubickyj et al. (2005): 6.07 Myrs) in contrast to 2.14 Myrs for nJ6.

The case nJ1a (thick lines) is identical to case nJ1, except that at $t=1.5$ Myrs, the solid accretion rate is artificially ramped down to zero on a short timescale as in P96's J1a model (see Fig. \ref{fig:J1J1a}, right panel). It shows another important effect (P96): The associated reduction of the core luminosity leads to an increase of the gas accretion rate due to the reduced thermal support of the envelope. Due to this, cross-over occurs at $t=3.38$ Myrs (P96: 3.32 Myrs). This effect can be important when two giant planets form concurrently, or when a planet migrates through a part of the disk which has previously been cleaned from planetesimals.

Numerical values for the most important quantities characterizing the nJ6, nJ1 and nJ1a models at particular moments in time are given in table \ref{tab:Jx}.  In this table we additionally list $\rho_{\textrm core}$ which is the mean density of the core.  We have found  in tests that at least for the nJ1 case, using a fixed core density (3.2 g/cm$^{3}$) instead of a variable one only plays a minor role. 
\begin{table}\footnotesize 
\caption{Models for the in-situ formation of Jupiter. }\label{tab:Jx}
\begin{center}
\begin{tabular*}{13cm}{llrrr}
\hline
Phase&  & nJ1 & nJ1a & nJ6 \\
\tableline
First luminosity peak & Time$^{a}$	 & 0.218 & 0.218  & 0.198  \\
                 & $M^{b}$ & 8.25& 8.25&8.30 \\
                           & $\mz^{\ b}$ &8.24 &8.24 & 8.28\\
                                    & $\mxy^{\ \ \  b}$&0.012 &0.012 & 0.024 \\
                                    & $\mdotz^{\ c}$ &1.01$\times$$10^{-4}$ & 1.01$\times$$10^{-4}$& 1.28$\times$$10^{-4}$ \\
                                     & $R/\rj^{\ d}$ & 40.6&40.6 & 40.8\\
                                     & $\rho_{\textrm core}^{\ \ \ \ \ e}$ & 2.99& 2.99 & 2.99\\
    				                 & $\log L^{f}$ & -4.97 & -4.97 & -4.86 \\ 
				  		      & $T^{g}$& 502.6& 502.6   & 214.1\\ \tableline
 							
 Cross over   point                & Time$^{a}$ & 6.62 & 3.38  & 2.14  \\
                        & $M^{b}$ & 33.22& 25.72&33.20 \\
                                  & $\mz$=$\mxy^{\ b}$ &16.61 & 12.86& 16.60\\
                                    & $\mdotz^{\ c}$ &2.67$\times$$10^{-6}$ & 0 & 5.85$\times$$10^{-6}$ \\
                                      & $\mdotxy^{\ \ \  c}$ &1.33$\times$$10^{-5}$& 2.14$\times$$10^{-5}$ &2.93$\times$$10^{-5}$ \\
                                     & $R/\rj^{\ d}$ & 128.2& 105.5  & 128.1\\
                                     & $\rho_{\textrm core}^{\ \ \ \ \ e}$ & 3.91& 3.64&  3.82\\
	 				             & $\log L^{f}$ & -6.21&-6.86 & -5.92\\
				                & $T^{g}$ &  201.1& 169.2 & 153.1\\ \tableline

 Onset of  limited gas     accretion                               & Time$^{a}$ & 7.09 & 3.59  & 2.39  \\
   ($\mdotxy=10^{-3}$ $\mearth$/yr)                       & $M^{b}$ & 68.99& 72.86& 65.34\\
                                                       & $\mz^{\ b}$ &21.01 & 12.86 &  20.66\\
                                                          & $\mxy^{\ \ \  b}$&47.98 &60.00 & 44.68 \\
                                                         & $\mdotz^{\ c}$ & 7.22$\times$$10^{-5}$ & 0 &  7.01$\times$$10^{-5}$\\
                                                          & $R/\rj^{\ d}$ & 217.5 & 226.8 & 209.6 \\
                                                          & $\rho_{\textrm core}^{\ \ \ \ \ e}$ & 4.27 & 3.65 & 4.14 \\
 						                           & $\log L^{f}$ & -4.81 & -5.07Ê& -4.81  \\ 
							                      & $T^{g}$ & 361.9  & 309.0 & 209.6\\ \tableline

Second luminosity  peak                                      & Time$^{a}$ & 7.31 & 3.82 &  2.61  \\
                                     & $M^{b}$ & 301.7  & 298.4 & 300.6 \\
                                              & $\mz^{\ b}$ &  34.27 & 12.86& 34.32 \\
                                                    & $\mxy^{\ \ \  b}$& 267.4 & 285.5   &266.3 \\
                                                     & $\mdotz^{\ c}$ &3.13$\times$$10^{-5}$   &  0  & 3.49$\times$$10^{-5}$\\
                                                      & $R/\rj^{\ d}$ & 2.41  & 1.95 &  2.38\\
                                                & $\rho_{\textrm core}^{\ \ \ \ \ e}$ &  7.21&   6.52 & 7.30 \\
	                      					 & $\log L^{f}$ &-3.34  & -3.35  &-3.29 \\ 
							             & $T^{g}$ & 1833  & 1262 & 1309\\ 
							              & $T_{\rm cent}^{\ \ \  g}$ & 6.65$\times$$10^{4}$  & 5.57$\times$$10^{4}$ & 6.62$\times$$10^{4}$ \\ \tableline

Present  day                
    (4.6 Gyr)                      & $M^{b}$ & 308.0& 305.6   & 307.3\\
                                   & $\mz^{\ b}$ & 34.43& 12.86 &34.55 \\
                                 & $\mxy^{\ \ \  b}$&273.6 & 292.7 &  272.8\\
                                      & $R/\rj^{\ d}$ & 1.05& 1.10  & 0.99\\
                                     & $\rho_{\textrm core}^{\ \ \ \ \ e}$ & 11.80 &  10.35& 12.10\\
 			         & $\log L^{f}$ &  -8.85 & -8.85  & -8.81 \\ 
			                  & $L/\lj^{\ i}$ &  1.61  &  1.63  & 1.78 \\ 
			                 & $T^{g}$ & 129.6  & 128.5 & 133.1\\ 
			                & $T_{\rm cent}^{\ \ \  g}$ & 2.35$\times$$10^{4}$  & 2.20$\times$$10^{4}$  & 1.73$\times$$10^{4}$\\ \tableline

\end{tabular*}
\begin{minipage}[l]{13cm}
$^{a}$: in Myrs. $^{b}$:in $\mearth$. $^{c}$: in $\mearth/yr$. $^{d}$: in Jovian equatorial radii $\rj=7.15\times10^{9}$ cm. $^{e}$: in g/cm$^{3}$. $^{f}$: in solar luminosities $\lsun$. $^{g}$: in K. $^{i}$: in present day Jovian intrinsic luminosity $\lj=8.7\times10^{-10}\lsun$.
\end{minipage}
\end{center}
\end{table}

\section{Luminosity evolution}\label{sect:lumievo}
Another observable quantity apart from the radius which is  obtained in the simulations is the planetary luminosity which is fundamental for another observational technique gaining recently importance as transits, namely the direct imaging technique. 

\begin{figure*}[h]
	      \centering
       \includegraphics[width=\textwidth]{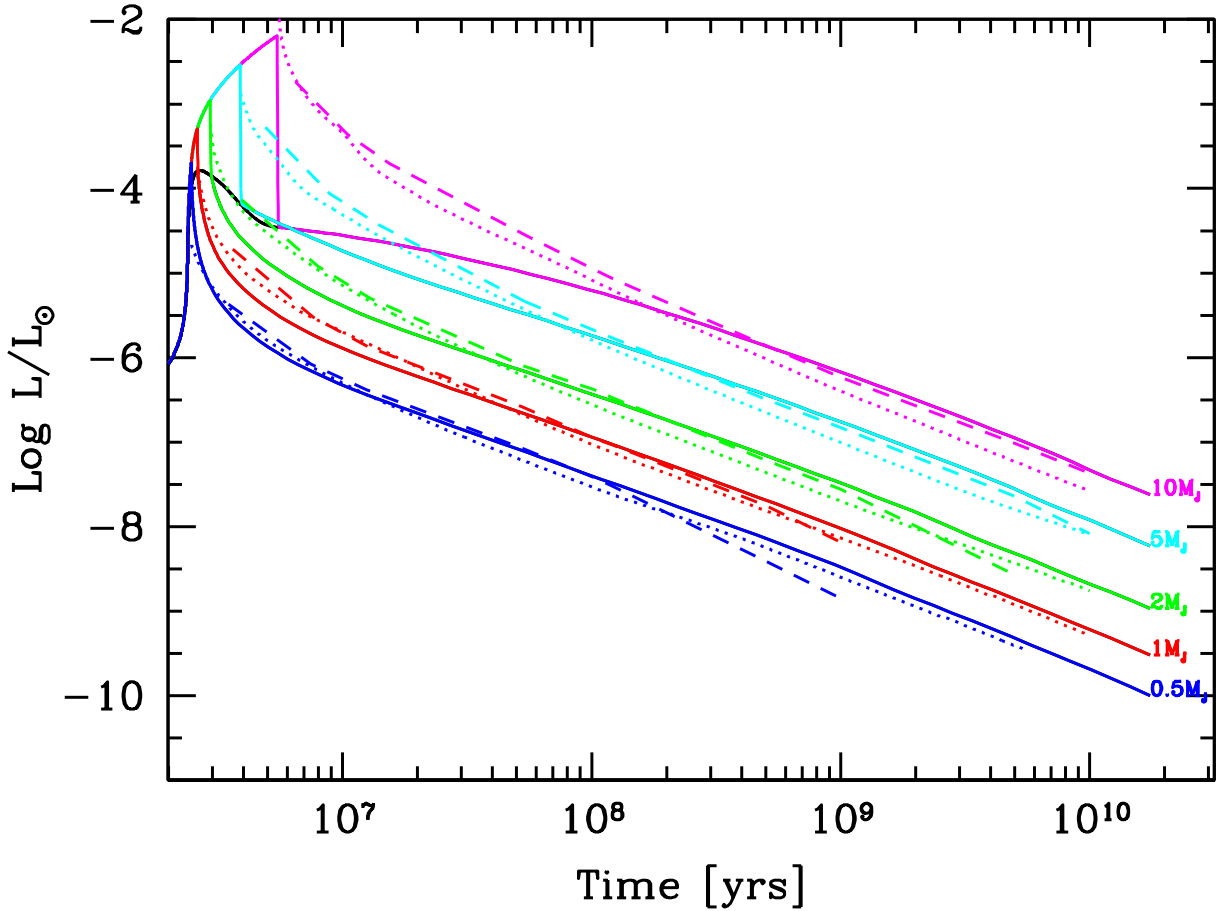}
       \caption{(Total) luminosity $L$ as a function of time, for planets with masses of 10, 5, 2, 1, and 0.5 $\mj$. For the  1 $\mj$ case, the line is the same as in Fig \ref{fig:J6lumi}.  Solid lines are results of this work while for comparison dashed lines are BC03 and dotted lines are BM97 models.  The solid black line additionally shows $L_{\rm int}$ during the detached phase.}\label{mordasinifig:Lcomp} 
\end{figure*}

Figure \ref{mordasinifig:Lcomp} shows the luminosity as a function of time for five planetary masses (10, 5, 2, 1, and 0.5 $\mj$). The formation phase for all five planets is identical to case nJ6, except that accretion was shut off at these masses. This means that the formation time increases with mass.

In the figure, we have also plotted the luminosity in the BM97  and BC03 models.  The moment corresponding to $t=0$ of these models has been set to the moment when accretion stops in our simulation. The sharp drop of $L$ particularly well visible for the 10 and 5 $\mj$ cases also corresponds to the end of accretion, when $L_{\rm ext}$ vanishes.
 
At late times, we see a generally good agreement. One notes that our models might have a certain tendency towards higher luminosities. This could be a consequence of the simple Eddington boundary conditions, in contrast to the atmospheric models used in BM97  and BC03.
The discrepancies are however not  larger than those mutually between BM97  and BC03.

At early times however, systematic discrepancies exist. The luminosities of the planets as found here are smaller than in the other two models. The strength of the difference increases with planetary mass. While no clear difference can be seen for the 0.5 $\mj$ case, and only a small one for the $1\mj$ case (see also Fig. \ref{fig:J6lumi}), it becomes very marked for the 5 and 10 $\mj$ case. Also the duration of the luminosity deficit increases with mass, and becomes as long as a few 100 Myrs for the most massive planet. 

This is of course nothing else than the well know difference between ``cold start'' and ``hot start'' models (Fortney et al. 2005, Marley et al. 2007) which is thus recovered in our simulations. There are however also differences to Marley et al. (2007): We note for example that in our work, more massive planets are also more luminous than smaller ones (except for a very short phase of the 10 $\mj$ planet), in contrast to  Marley et al. (2007). It also seems that the luminosity deficit is in general less strong in our simulations, so that one could talk of ``warm start'' models.  This will be further studied in future work. 

As  Marley et al. (2007) we stress that these results depend on the details of the accretion process, which we have described in a strongly simplified way, giving rise to significant uncertainy. But it confirms that the  luminosity  especially of massive young giant planets is currently poorly constrained. 

\section{Synthetic mass-radius diagram} 
With the formation model presented in section  \ref{sect:model} it is possible to obtain a synthetic mass-radius diagram by planetary population synthesis (Ida \& Lin 2004, Mordasini et al. 2009a). 

The probability distributions for the Monte Carlo variables, as well as the synthesis parameters like for example the $\alpha$ parameter for the disk viscosity, or the  efficiency factor for type I migration are the same as in Mordasini et al. (2009a). The initial gas surface density profile however is now taken to be $\Sigma(r,t=0)=\Sigma_0\left(r/5.2\, \mathrm{AU}\right)^{-0.9}\exp(-\left(r/30\,  \mathrm{AU}\right)^{1.1})$ as indicated by the observation of young circumstellar disks (Andrews et al. 2009). The stellar mass if fixed to 1 $\msun$.

\begin{figure*}[t!]
\begin{minipage}{0.5\textwidth}
	      \centering
       \includegraphics[width=\textwidth]{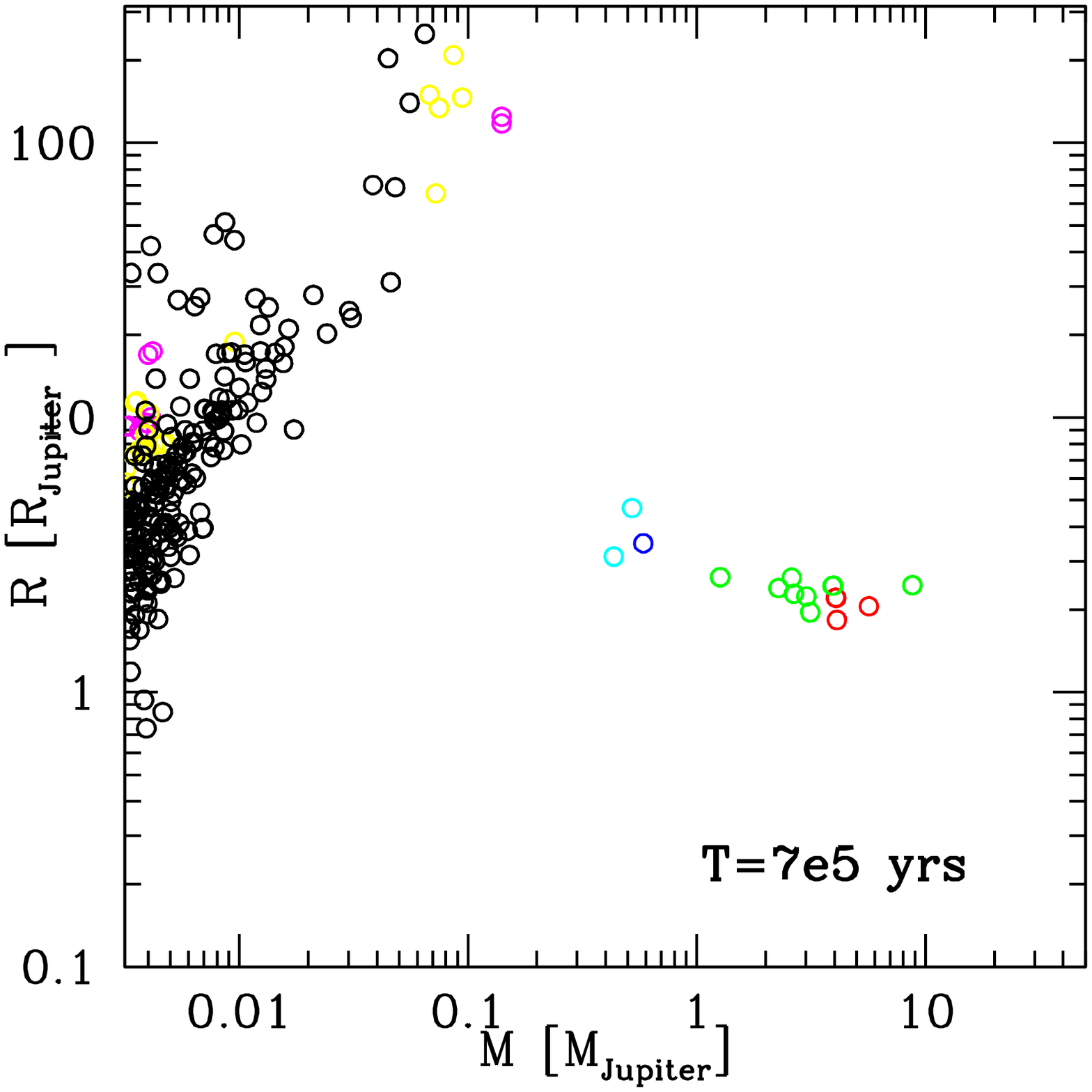}
     \end{minipage}\hfill
     \begin{minipage}{0.5\textwidth}
      \centering
       \includegraphics[width=\textwidth]{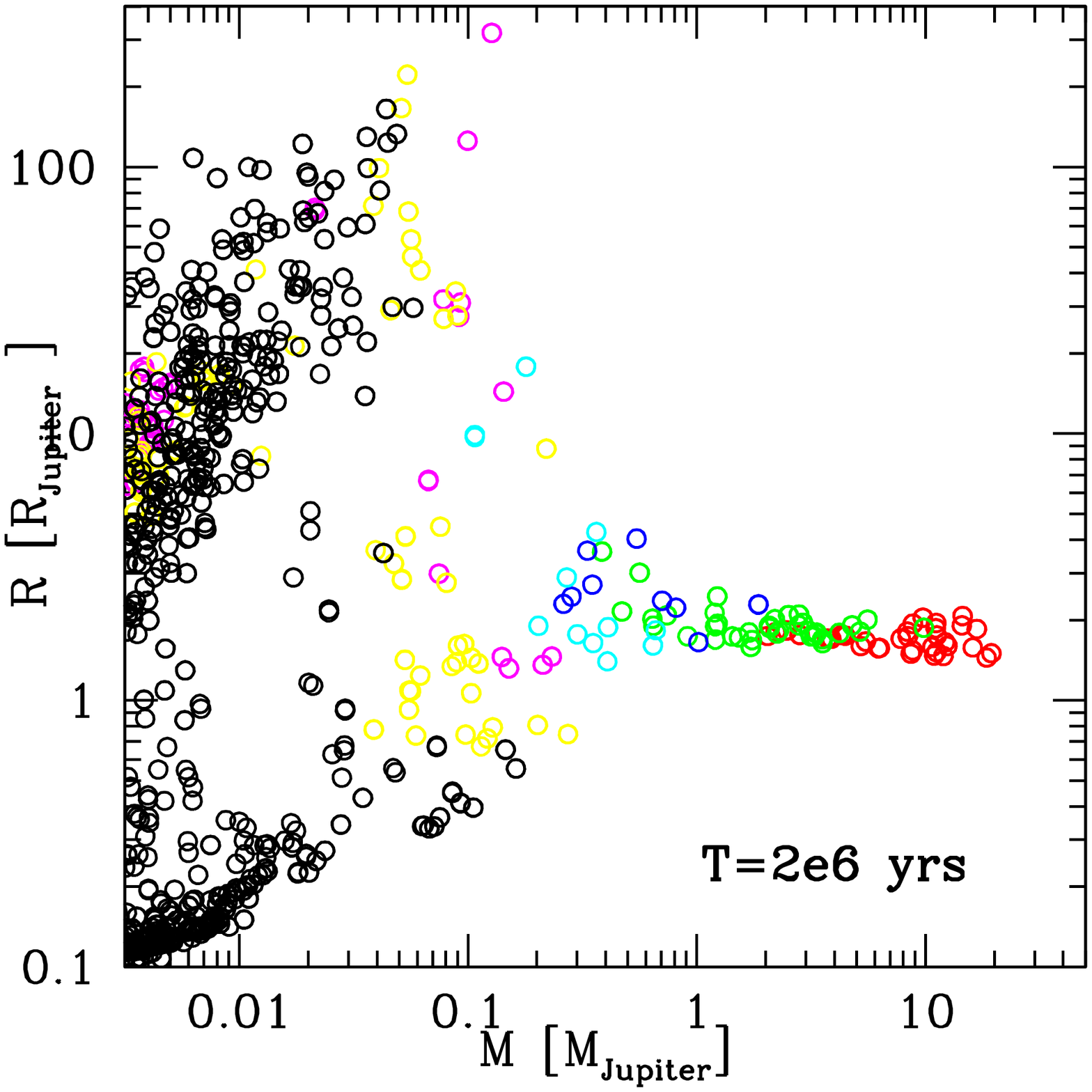}
     \end{minipage}
     \begin{minipage}{0.5\textwidth}
	      \centering
       \includegraphics[width=\textwidth]{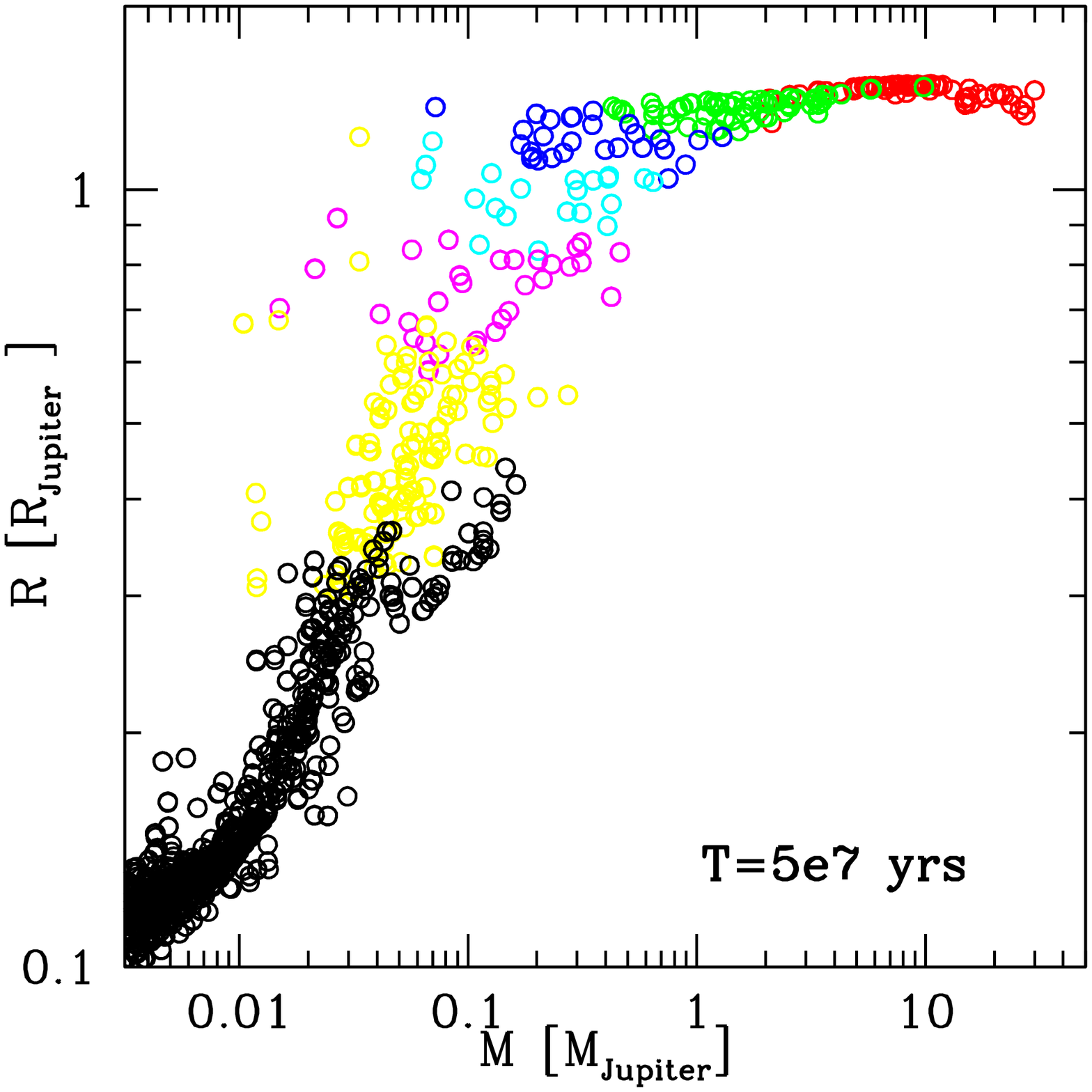}
     \end{minipage}\hfill
     \begin{minipage}{0.5\textwidth}
      \centering
       \includegraphics[width=\textwidth]{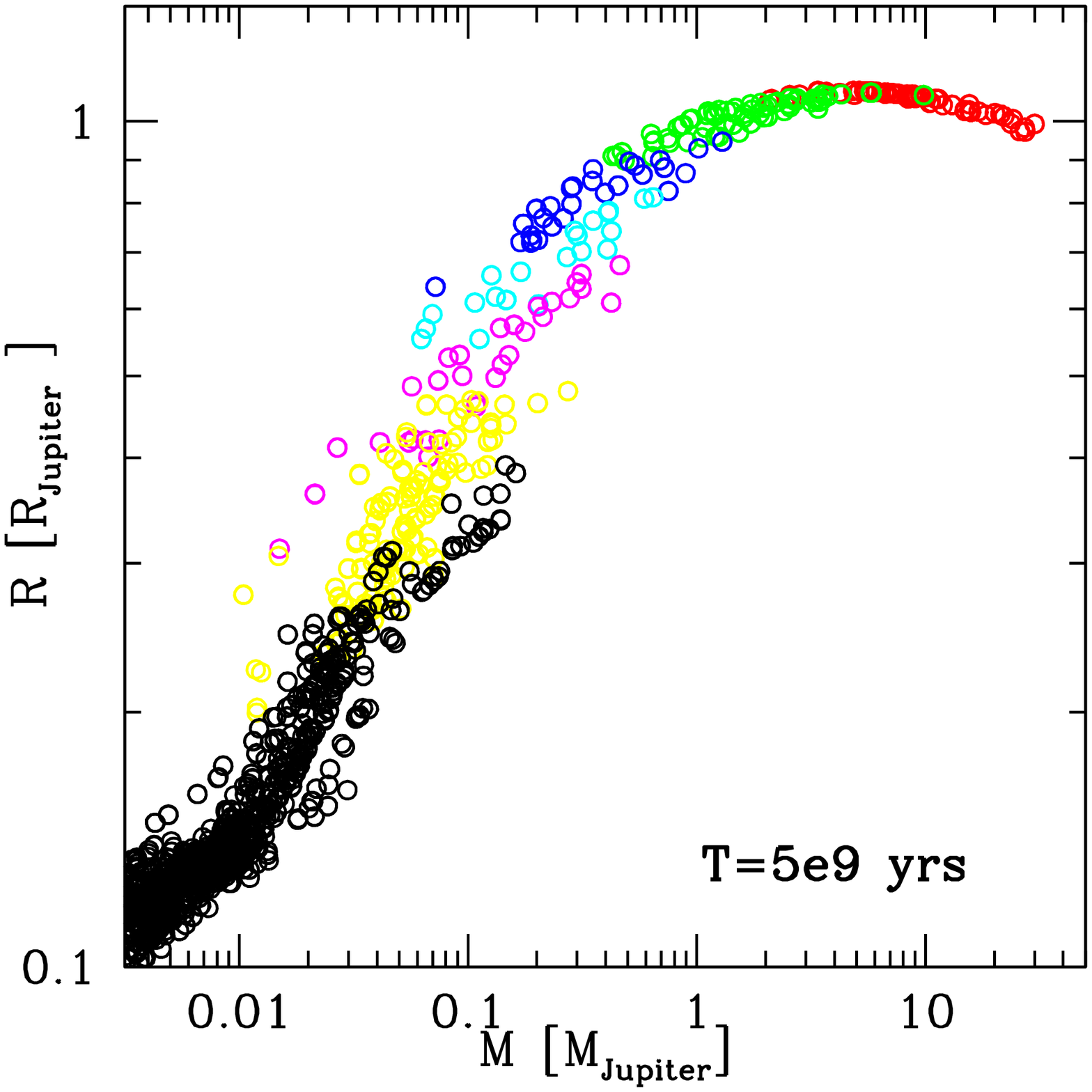}
     \end{minipage}
     \caption{Temporal evolution of the mass-radius diagram for synthetic planets. The minimal mass corresponds to $1\mearth$. Four moments in time are shown, as indicated in the panels. The colors show the fraction of heavy elements in the planet. Red: $Z$$<$5\%. Green: 5$<$$Z$$<$20\%. Blue:  20$<$$Z$$<$40\%. Cyan: 40$<$$Z$$<$60\%. Magenta: 60$<$$Z$$<$80\%. Yellow: 80$<$$Z$$<$95\%. Black: $Z>$95\%. }\label{fig:formationmr} 
\end{figure*}

Figure \ref{fig:formationmr} shows the temporal evolution of the synthetic mass-radius diagram in four panels. The color coding indicates the fraction of heavy elements $Z=\mz/M$. We assume for simplicity in all simulations that all heavy elements reside in the core, so that $\mz=\mcore$.

Note that some low mass planets at smaller distances from the star were excluded from the figures shown here, as they start to inflate and overflow their Hill sphere once the external nebula is removed. This means that they would start to loose mass, an effect that we currently do not include in the model. Note further that we currently neither check whether the primordial H$_{2}$/He atmosphere of a (low mass) planet is stable against evaporation. Both points will be addressed in future work.  

The top left panels shows the $M-R$ diagram in an early stage of the formation phase at $t=7\times10^{5}$ years. The first giant planets have already formed which is possible in metal rich, massive disks. One can distinguish two groups in the diagram: Low mass, pre-collapse (and pre-runaway) planets which have a very large radius ($\sim R_{\rm H}$), and higher mass post-collapse planets which are in runaway gas accretion, with a radius of roughly 2 $\rj$. One notes that the transition from one group to the other occurs at a mass of about $0.2\mj\approx60\mearth$, as we have seen for the Jupiter simulations. The colors show that the core and the envelope have at this point about the same mass, again as seen in the Jupiter models. The two groups are separated: This is a consequence of the relatively short duration of the collapse.  

The top right panel shows the situation at 2 Myrs. This corresponds to the middle of the formation phase. One sees that there are now also low mass planets with small radii. This are planets which never undergo a phase of  runaway gas accretion but collapse after the protoplanetary disk has disappeared. For the giant planets we see that for one mass, there is a significant spread in associated radius. This is a visible imprint of different formation histories.

The bottom left panel at $t=50$ Myrs shows the early stage of the evolution phase. Mass accretion is now over for about 30 Myrs, and all planets have collapsed, so that the planets only slowly cool and shrink at constant mass. The radius of typical giant planets is now between 1.2 and 1.4 $\rj$. There is still a larger spread in $R$ for giant planets of nearly identical masses than later on. This is because the formation history still plays a role.

The bottom right panel finally shows mature planets at an age of $5\times10^{9}$ years. For  giant planets we see the characteristic maximum of the radius (found here to occur at about $4\mj$) which is due to the change of the degree of degeneracy in the interior (e.g. Chabrier et al. 2009). We also note that at given mass, planets with a higher Z have a smaller radius, as it must be (e.g. Fortney et al. 2007).

Concerning the overall shape, we see that low mass planets have small radii, as they consist mostly of solids, while high mass planets have large radii, as they consist mostly of gas, as indicated by the color coding showing the decrease of $Z$ with $M$. This is a consequence of the fact that low mass cores cannot accrete large amounts of gas as their Kelvin-Helmholtz timescale is too long. Such planets never undergo gas runaway accretion. In a symmetrically way to this must massive cores always accrete significant amounts of gas (at least if they form during the lifetime of the  gaseous disk) as they must trigger the collapse of the envelope. Neither a (nearly) core-free e.g. 10$\mearth$ planet (with $R\sim1\rj$ at 1 AU, Fortney et al. 2007), nor a pure ice $3\mj$  planet (with $R\sim0.4\rj$, Charbonneau et al. 2010) is formed. We see that the synthetic $M-R$ relationship contains the imprint of the basic concepts of the core accretion formation theory, and that the (observed) $M-R$ relation thus constrains the exact mechanism of this theory. 

\subsection{Comparison with observation}
\begin{figure*}[h!]
       \includegraphics[width=\textwidth]{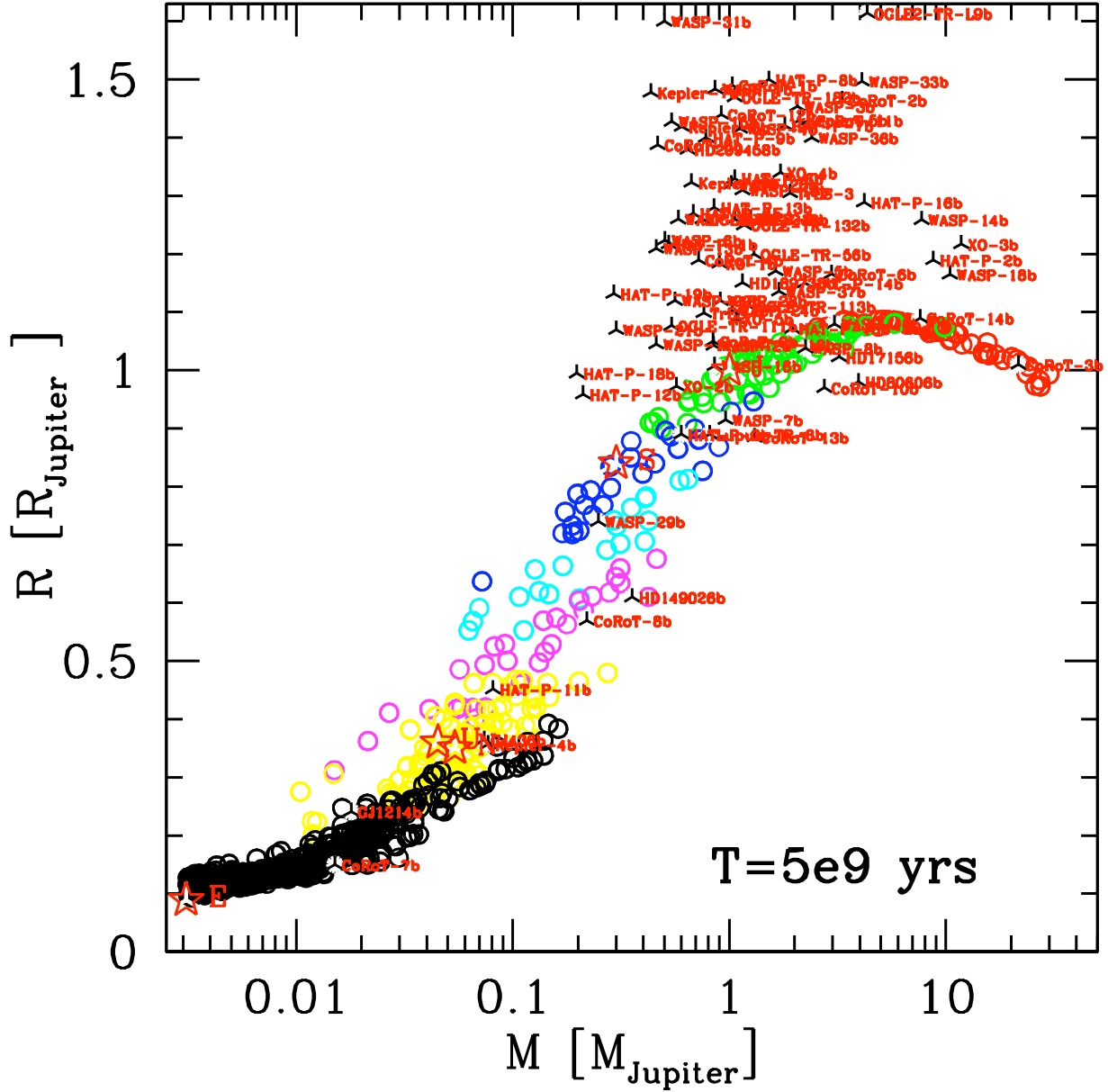}
      \centering
       \caption{Mass-radius diagram for synthetic planets at an age of $5\times10^{9}$ yrs (same as Fig. \ref{fig:formationmr}, bottom right) compared to the planets of the solar system (large symbols labelled E, J, S, U, N)  and to transiting extrasolar planets. The colors again shows for the synthetic planets the fraction Z of heavy elements, as in Fig. \ref{fig:formationmr}.  }\label{fig:mr}  
\end{figure*}

Figure \ref{fig:mr} again shows the synthetic $M-R$ diagram at 5 Gyrs (but now in linear units for $R$), together with the planets of the solar system, and transiting extrasolar planets with $R<1.6\rj$.  

Regarding the comparison with the planets of the solar system we note a good agreement for the gas giants and the ice giants both in terms of their position in the $R-M$ diagram, as well as in terms of the bulk composition as indicated by the color code (Jupiter: 3$<$Z$<$13\%, Saturn: 21$<$Z$<$31\%, Ice giants: 85$<$Z$<$95\%, Guillot \& Gautier 2007). It is clear that the Earth does not posses a primordial H$_{2}$/He atmosphere, as assumed here. But if one compares with the synthetic core radii, then is this planet also well reproduced.

When making the comparison with the extrasolar planets one should bear in mind that the synthetic planets have all semimajor axes larger than 0.1 AU, and that no extra mechanism which could cause inflation is included here. It is therefore no surprise to see that the transiting planets which have radii larger than about 1.1$\rj$ cannot be reproduced. 

One however also sees that HD\,80606b (Naef et al. 2001, Moutou et al. 2009) and COROT-10b (Bonomo et al. 2010)  have a radius that is smaller than found in the synthetic population.  The radius discrepancy cannot be explained by a higher age of the stars alone (7.3 Gyr for HD\,80606, 1-3 Gyr for COROT-10). The high densities of these planets have been noted previously  (Moutou et al. 2009, Pont et al. 2009, Bonomo et al. 2010).  This means that these planets contain a higher amount than the synthetic planets, where the overall highest core masses are about 180 $\mearth$. An interesting point to note is that both planets are quite peculiar in terms of their orbital characteristics, too: Both are found on very eccentric orbits (compared to other planets with a similar orbital period, see Bonomo et al. 2010), and HD\,80606b additionally has a clear spin-orbit misalignment. These factors could indicate that these planets underwent  a giant impact with another planet (see Baraffe et al. 2008). 

For transiting exoplanets of a lower mass we see a good agreement between predicted and observed M-R loci, including planets as different as Corot-7b (L{\'e}ger et al. 2009) or GJ\,1214b (Charbonneau et al. 2009).  It is interesting to note that intermediate mass planets (with a mass say between Neptune and Saturn), can have for a given mass a large spread of more than a factor 2 for the radius. This reflects the diversity of their heavy element contents, as shown by the colors.

\begin{figure*}
\begin{minipage}{0.5\textwidth}
	      \centering
       \includegraphics[width=\textwidth]{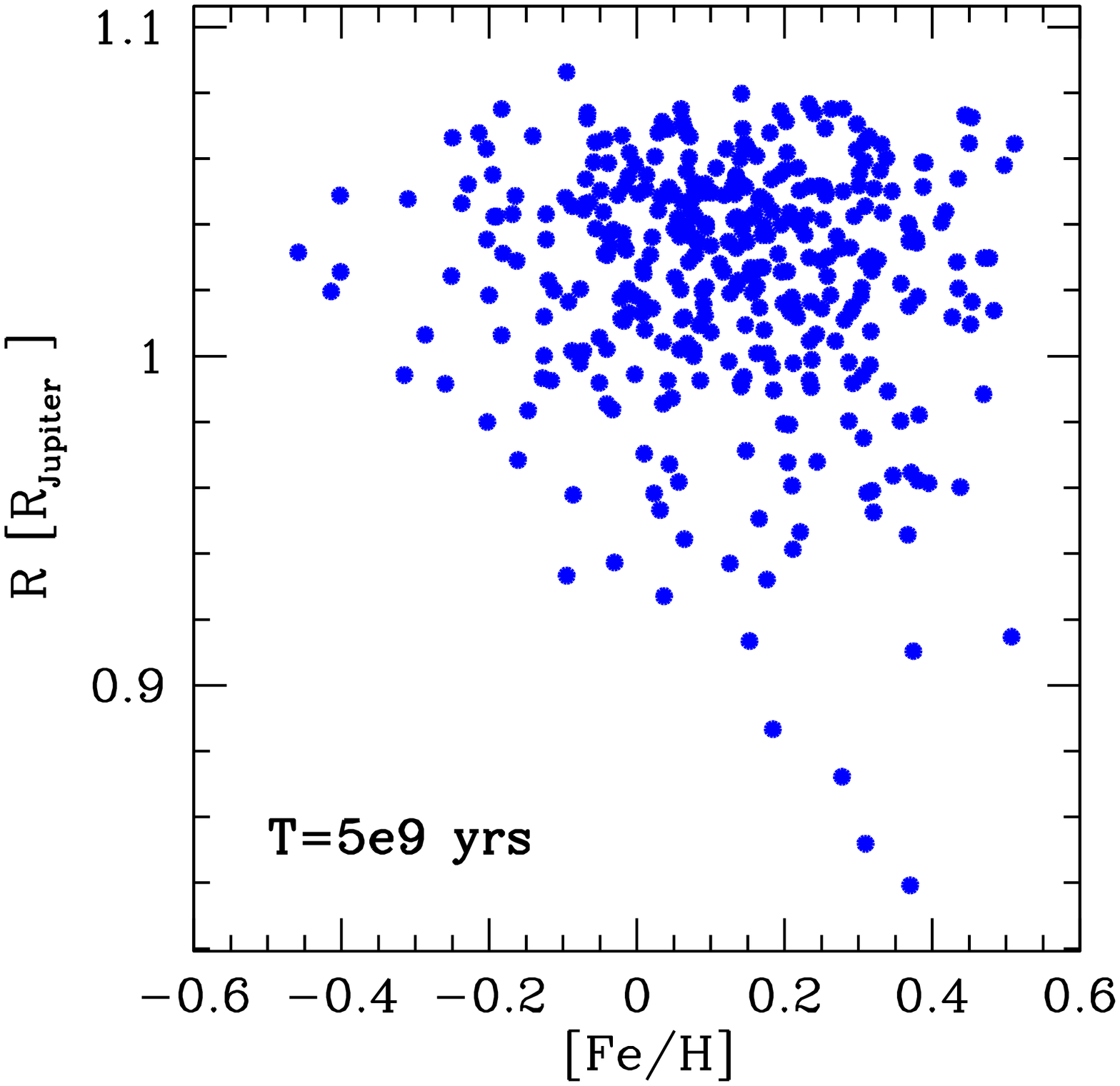}
     \end{minipage}\hfill
     \begin{minipage}{0.5\textwidth}
      \centering
             \includegraphics[width=\textwidth]{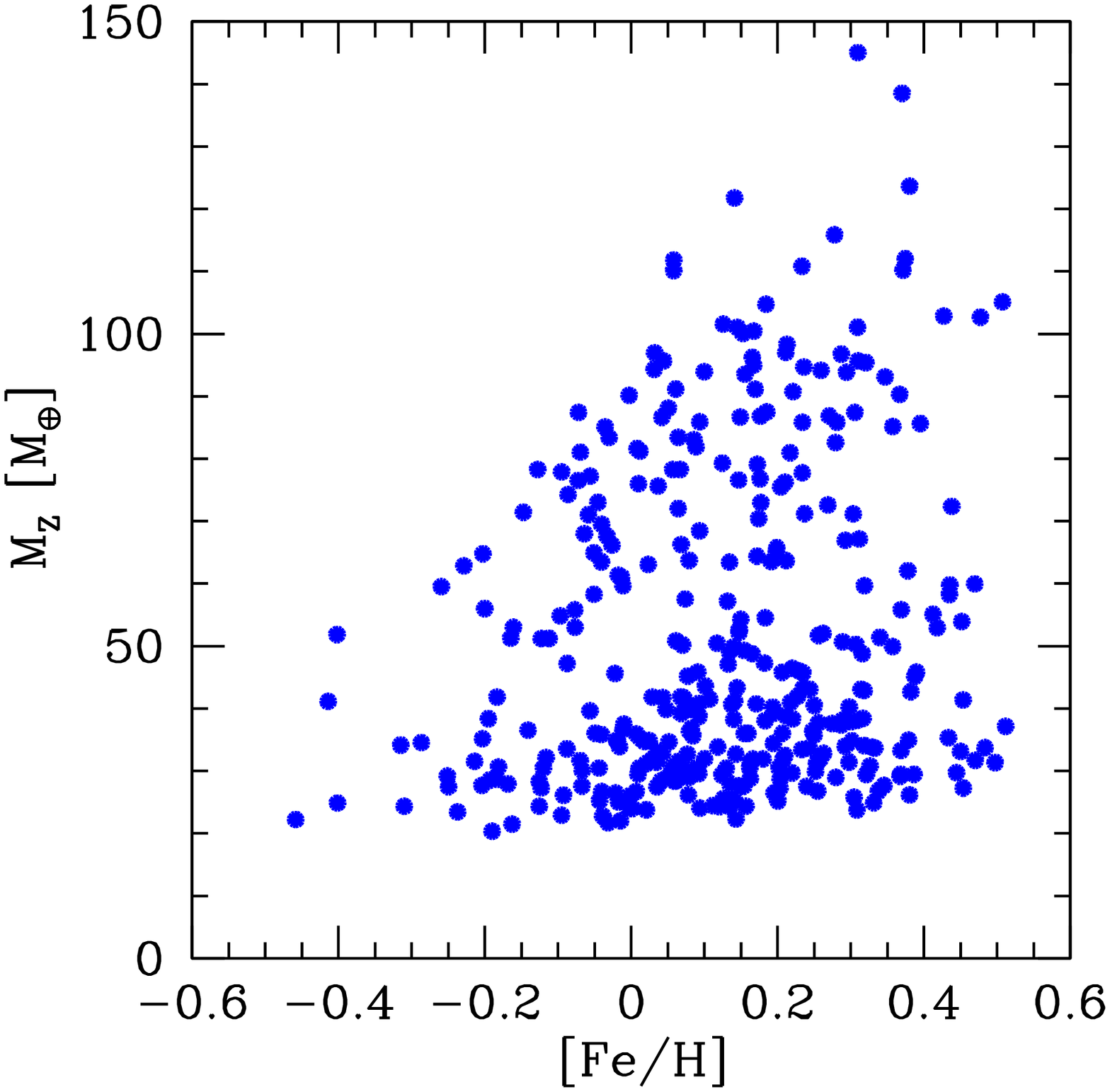}
     \end{minipage}
        \caption{Left panel: Planetary radius at 5 Gyrs for synthetic planets with a mass of 1 to 3 $\mj$ as a function of stellar/disk [Fe/H]. Right panel: Mass of  heavy elements $\mz$ inside the same synthetic planets, again as a function of [Fe/H]. }\label{mordasinifig:fehR} 
\end{figure*}

Models of Guillot et al. (2006) and Burrows et al. (2007) suggest that planets with a high amount of heavy elements tend to be found around stars with a high [Fe/H], a correlation  which was reproduced in the synthetic population of Mordasini et al. (2009b). Figure \ref{mordasinifig:fehR}, left panel, shows the radius of synthetic planets at 5 Gyrs with a mass between 1 and 3 $\mj$ as a function of the stellar (and thus disk, see Mordasini et al. 2009a) [Fe/H]. 

One notes that the minimal radius decreases with increasing stellar metallicity. This is because the maximal core mass increases with [Fe/H], as can be seen in the right panel. The maximal radius however hardly changes with metallicity, as all planets contain at least 20 $\mearth$ of heavy elements, which is necessary to trigger gas runaway accretion, and this number nearly does not increase with [Fe/H]. This is because also at high [Fe/H], low absolute amounts of solids can exist in a disk, if the gas mass of the disk is small. On the other hand, there are no planets with a small $R$ at low [Fe/H], as metal poor disks do not dispose of the necessary high quantities of metals for incorporation into the planets. 

\section{Conclusions}
We have presented an updated planet formation code which allows to calculate in a self-consistent way planetary radii and luminosities during the entire formation and evolution phase. We have then studied with it the formation of Jupiter in the in-situ approximation, and compared the results to other works, finding good agreement during the formation phase. Even though clearly simplified compared to other evolutionary models, we have found reasonable  agreement also for the long term evolution, keeping in mind our purpose of planetary population synthesis. We have for example also found the discrepancy between ``hot start'' and ``cold start'' models.  The fundamental final properties of the model Jupiter are in agreement with observational data.

Finally, we have calculated a synthetic planetary population around  solar-like stars and presented in this way how the planetary mass-radius diagram comes into existence. We show that the general shape of the observed and the synthetic mass-radius relationship is in good agreement. The exception are bloated Hot Jupiters which can only be explained with additional mechanism which are related to their proximity to the host star.  The general shape of the $M-R$ relation can be understood by a decrease of the heavy element fraction with mass. This is a direct consequence of the core accretion theory: Low mass planets cannot accrete massive gas envelopes because of their long cooling timescale. Massive cores ($\sim20\mearth$) on the other hand trigger gas runaway accretion. The spread of planetary radii for one given mass can however by substantial due to different formation histories. We thus conclude that the observed $M-R$ diagram carries the imprint of the planetary formation mechanism.

\acknowledgments{We thank K.-M. Dittkrist, C. Ormel, C. Broeg, J. Leconte, I. Baraffe and G. Chabrier for useful discussions. This work was supported in part by the Swiss National Science Foundation. Christoph Mordasini acknowledges the support as a fellow of the Alexander von Humboldt foundation, and the hospitality of the Kavli Institute for Theoretical Physics at UCSB in spring 2010 (funded by the U. S. NSF through Grant PHY05-51164). Yann Alibert is thankful for the support by the European Research Council under grant 239605.  }

\end{document}